\documentclass[10pt,letterpaper]{article}
\usepackage{opex3}
\usepackage{amsmath}
\usepackage{cite}
\usepackage{subfigure}
\begin{document}
\title{A collinear, two-color optical Kerr effect shutter for ultrafast time-resolved imaging}
\author{Harsh Purwar,$^{1,*}$ Sa\"id Idlahcen,$^{1}$ Claude Roz\'e,$^{1}$ David Sedarsky,$^{2}$ and Jean-Bernard Blaisot$^{1}$}
\address{$^{1}$CORIA-UMR 6614-Normandie Universit\'e, CNRS-Universit\'e et INSA de Rouen, Campus Universitaire du Madrillet, 76800 Saint Etienne du Rouvray, France\\$^{2}$Applied Mechanics, Chalmers University of Technology, H\"orsalsv\"agen 7B, 41296 G\"oteborg, Sweden}
\email{$^{*}$harsh.purwar@coria.fr}
\begin{abstract}
Imaging with ultrashort exposure times is generally achieved with a crossed-beam geometry. In the usual arrangement, an off-axis gating pulse induces birefringence in a medium exhibiting a strong Kerr response (commonly carbon disulfide) which is followed by a polarizer aligned to fully attenuate the on-axis imaging beam. By properly timing the gate pulse, imaging light experiences a polarization change allowing time-dependent transmission through the polarizer to form an ultrashort image. The crossed-beam system is effective in generating short gate times, however, signal transmission through the system is complicated by the crossing angle of the gate and imaging beams. This work presents a robust ultrafast time-gated imaging scheme based on a combination of type-I frequency doubling and a collinear optical arrangement in carbon disulfide. We discuss spatial effects arising from crossed-beam Kerr gating, and examine the imaging spatial resolution and transmission timing affected by collinear activation of the Kerr medium, which eliminates crossing angle spatial effects and produces gate times on the order of $1$~ps. In addition, the collinear, two-color system is applied to image structure in an optical fiber and a gasoline fuel spray, in order to demonstrate image formation utilizing ballistic or refracted light, selected on the basis of its transmission time.
\end{abstract}


\section{Introduction}
The non-intrusive nature of optical diagnostics make them essential tools for the study of physical systems, which are often easily perturbed. However, many real-world applications and phenomena of interest are intrinsically linked to turbid environments where light scattering and attenuation strongly limit the efficacy of optical methods. In a wide range of applications, from imaging in biological tissues \cite{Alfano1997,Wang1991a}, to measurements in turbulent multiphase flows \cite{Linne2013a}, key information is scrambled by the distortion imparted to the light signal as it transits the measurement volume \cite{Berrocal2007}. Informative optical diagnostics in such media require detailed understanding of the light source, propagation and scattering in the measurement volume, and a detection arrangement tailored to select the meaningful parts of the transmitted light signal.

To this end, ultrafast time-gating can provide an effective means of segregating high integrity portions of the collected signal from light disturbed by scattering interactions. On average, photons which participate in more interaction events traverse a more circuitous path through the medium and are statistically more likely to be distorted or redirected from their original trajectories. This difference in optical path length results in a temporal spreading of the light intensity such that heavily distorted signal components arrive at later times. Time gating (time-based filtering) allows selection of the optical signal within a short temporal window and can be applied to separate the light that is scrambled due to multiple interactions from the optical signal that retains high-fidelity information of the object characteristics \cite{Calba2008}.

However, this requires that the time window of the gating must be short -- on the order of $10$'s of picoseconds, or shorter. Assuming a $100$~fs input pulse and a $\sim1$~cm turbid measurement volume, the typical transmitted pulse duration will be on the order of $50$-$100$~ps, with most of the informative signal present in the first $500$ fs of the transmitted signal \cite{Berrocal2009}. 

Practical electronic modulation is generally limited to nanosecond timescales, though advances in piezo-electronic devices have shown that it may be possible to gate at a few hundreds of picoseconds in specialized applications \cite{Newns2012}. Thus, even the fastest electronically controlled gating schemes fall short of the performance demanded by most time-resolved imaging applications in turbid media.

\section{Picosecond time-gated imaging}
\begin{figure}[htbp]
\centering
\subfigure[Optical gate in close state]{\includegraphics[width=0.4\columnwidth]{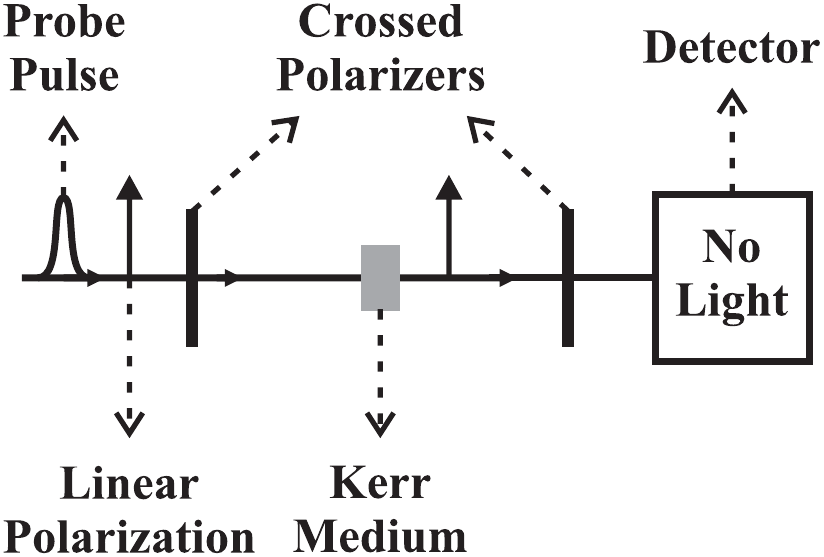}}\quad\quad
\subfigure[Optical gate in open state]{\includegraphics[width=0.4\columnwidth]{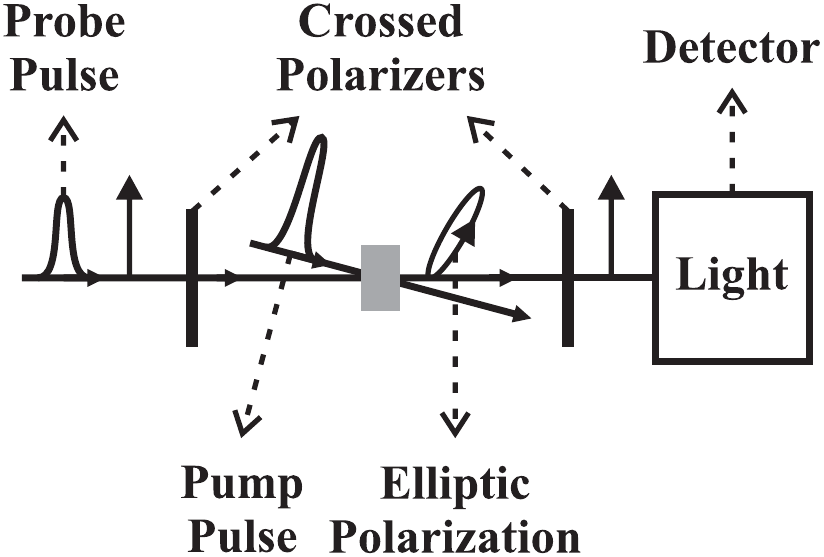}}
\caption{Schematic showing the working principle of crossed-beam OKE based time-gating.}
\label{non-coll-cartoon}
\end{figure}

Over the past two decades, a number of nonlinear optical processes have been exploited to form sub-picosecond gating mechanisms, including stimulated Raman \cite{Duncan1992}, second harmonic generation (SHG) \cite{Yoo1991,Idlahcen2011}, and the optical Kerr effect (OKE) \cite{Wang1991}. In terms of imaging applications, the latter approach remains the most popular in the literature and has generated renewed interest for imaging high-speed transient phenomena in scattering environments \cite{Linne2013a}. OKE gating arrangements typically achieve precise timing control by splitting the energy of an ultra-short laser pulse and adjusting the path length difference between the two beams in a pump/probe scheme to control the nonlinear interaction. The probe, or imaging pulse is used to illuminate the volume of interest, and the pump pulse is delayed and used to drive the nonlinear optical process for light selection. Thus, the temporal width of the gate window depends on the pump pulse duration and the relaxation time of the nonlinear optical effect. Figure~\ref{non-coll-cartoon} illustrates the working principle of such a pump-probe arrangement used to form an OKE time-gate using the commonly employed crossed-beam geometry. Since OKE time-gating is the focus of the present work, the operational details of this typical arrangement will be discussed in slightly more detail.

\begin{figure}[htbp]
\centering
\includegraphics[width=0.3\columnwidth]{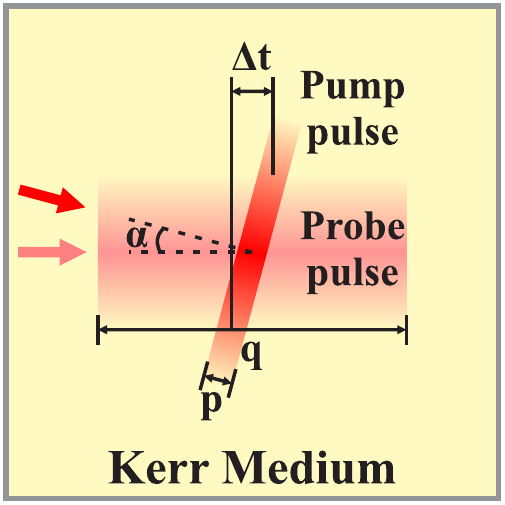}
\caption{Schematic showing the overlap of pump and imaging pulses inside the Kerr medium in a crossed-beam configuration. $\alpha$: angle between pump and probe beams, $p$: pump pulse width or duration, $q$: probe pulse width or duration (extended due to its interaction with the object/sample), $\Delta t$: time difference between the two sides of the obtained images.}
\label{Kerr_overlap}
\end{figure}
\begin{figure}[htbp]
\centering
\includegraphics[width=0.8\columnwidth]{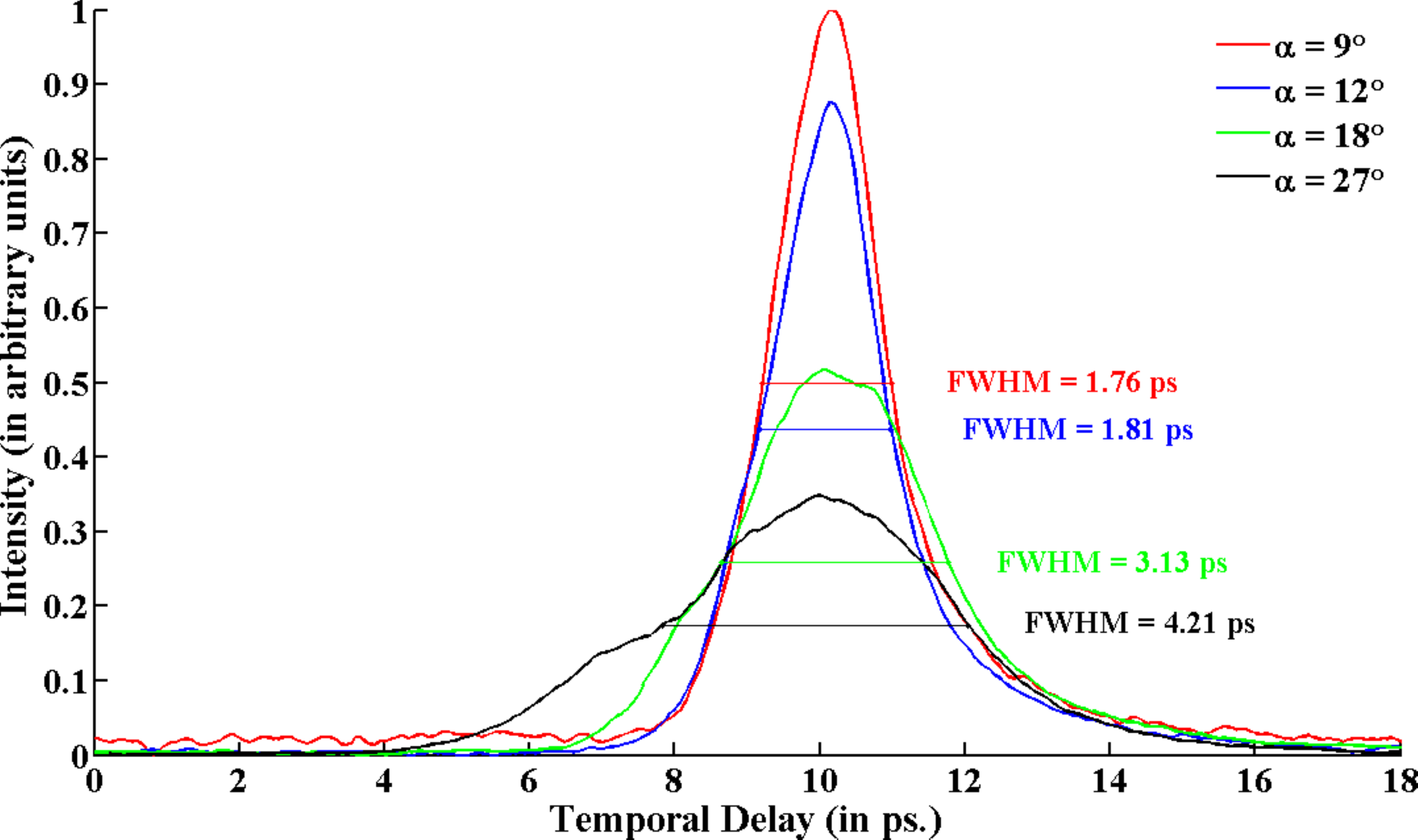}
\caption{Temporal profiles for the crossed-beam configuration of the optical gate with $1.0$~mm of CS$_2$ as Kerr medium for different crossing angles ($\alpha$) between the pump and the probe beams.}
\label{angle_delay}
\end{figure}

In the crossed-beam OKE gating arrangement, an off-axis gating pulse induces a time-dependent birefringence in a medium exhibiting a strong OKE response (commonly carbon disulfide) which is coupled with a polarizer positioned to fully attenuate the linearly polarized imaging beam which is aligned to propagate along the optical axis of the system. By setting the delay of the gate pulse, a selected portion of the imaging light within a short time window experiences a polarization change which allows this selected light to transit the polarizer to form an ultrashort image.  Here, the polarization change experienced by the image pulse is contingent on the precise overlap of the pump and probe pulses in the Kerr medium. In this crossed-beam configuration, perfect spatial and temporal overlap of the pulses is not possible; due to the crossing angle of the beams the temporal selection carried out by the pump pulse varies across the width of the imaging beam (see Fig.~\ref{Kerr_overlap}). When the time-gate duration is on the order of a few picoseconds this spatial error is generally small, but the error becomes more significant for shorter gate times, larger imaging beam widths, or steeper crossing angles. Figure~\ref{angle_delay} shows the temporal profiles for the CS$_2$ based optical time-gate in this arrangement for different crossing angles between the pump and the probe beams.

One approach for mitigating the spatio-temporal variation across the image beam is to reduce the crossing angle of the pump and image beam paths. Unfortunately , this increases forward-scattering of the pump pulse energy into the imaging path, and it becomes necessary to employ some additional scheme to separate the weak imaging light from the intense background. Extending this approach to its logical conclusion, if one is able to separate the pump and imaging pulses by some other means, the spatial overlap of the pulses can be maximized by reducing the crossing angle to zero.

This work presents a collinear time-gated OKE imaging arrangement in carbon disulfide (CS$_2$), where type-I frequency doubling is employed to spectrally differentiate the imaging light from the pump pulse energy. The following sections discuss the imaging spatial resolution achieved by the collinear OKE system and the light transmission timing resulting from collinear activation of the Kerr medium. This system is then applied to two imaging situations where the OKE gate is adjusted to capture refracted and ballistic light images in order to demonstrate the light selection capabilities of the two-color system.

\section{Collinear OKE time-gating}
\begin{figure}[htbp]
\centering
\includegraphics[width=0.9\columnwidth]{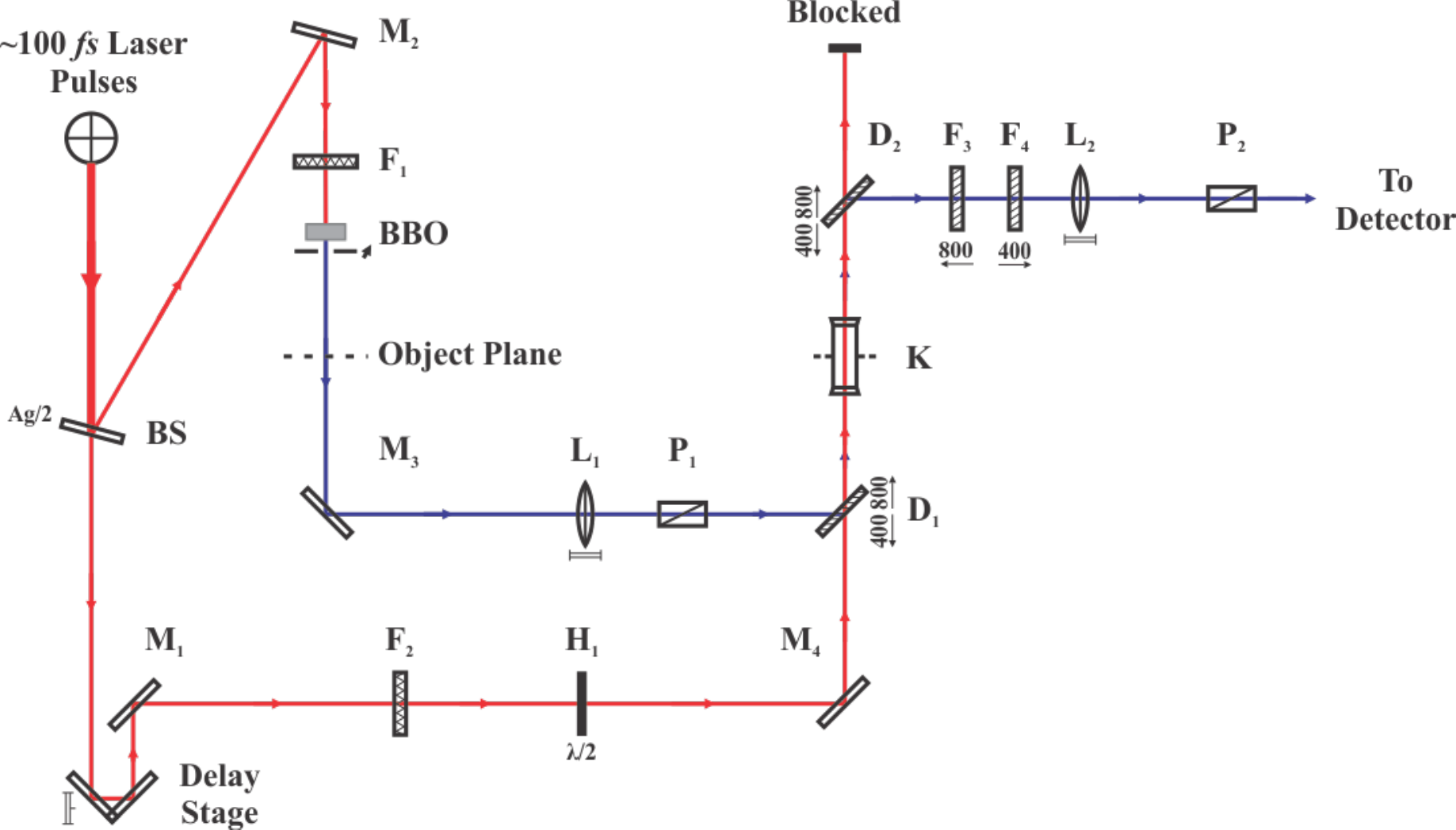}
\caption{Schematic of the experimental set-up for dual wavelength OKE based time-gate with collinear incidence of pump and probe beams at the Kerr medium. $BS$: 50/50 beamsplitter, ($M_1$, $M_2$, $M_3$, $M_4$): mirrors, ($D_1$, $D_2$): dichroic mirrors, ($P_1$, $P_2$): polarizers, $H_1$: half waveplate, ($F_1$, $F_2$): neutral density filters, $BBO$: $\beta$-barium borate crystal, ($L_1$, $L_2$): bi-convex lenses with focal lengths $300$ mm and $100$ mm respectively, $F_3, F_4$: filters to block pump ($800$~nm) and select only probe ($400$~nm) for detection.}
\label{dcc_setup}
\end{figure}

Figure~\ref{dcc_setup} shows a schematic of the collinearly pumped OKE time-gated imaging system used in this work. The femtosecond laser pulse used to generate the image and pump pulses is supplied by a regenerative amplifier (Coherent Libra) seeded by a mode-locked Titanium-Sapphire oscillator (Coherent Vitesse). This system generates laser pulses with pulse width $\sim100$~fs centered at wavelength of $800$~nm and a peak power of $35$~GW at a $1$~kHz repetition rate. Each laser pulse from the amplifier is directed into the optical system where it is split into pump and imaging beam paths by a $50$:$50$ beamsplitter.

The imaging pulse passes through an ND filter, and is frequency doubled by means of type-I second harmonic generation (SHG) in a beta-barium borate (BBO) crystal to produce imaging light with a center wavelength of 400 nm. The conversion efficiency of the SHG process is on the order of $30\%$. After interaction with the measurement volume the imaging pulse is directed through a polarizer, and on to the CS$_2$ cell by a dichroic mirror. This dichroic element transmits the $800$~nm pump pulse while reflecting the $400$~nm imaging pulse, effectively combining the image and pump beam paths to enable collinear propagation into the Kerr medium. Here, the collinear arrangement was selected to allow optimal spatial overlap of the pulses within the CS$_2$.

The Kerr medium (CS$_2$) was placed at the image plane of the lens $L_1$ so that the spatial frequencies are gated equally through the optical gate, thereby producing an image with better resolution compared to the case where the Kerr medium is place at the Fourier plane of $L_1$. Although, this requires the switching or pump beam to be more homogeneous and large enough to encompass the image formed by $L_1$ at its image plane. This can be achieved easily in this collinear configuration since the temporal characteristics of the optical gate do not depend on the spatial size of the switching beam. In case the switching beam is not homogeneous throughout in space, the resulting time-gated images would also not be transmitted homogeneously through the optical gate.

The pump pulse is directed from the beamsplitter through a computer controlled delay stage with a minimum step-size of $1$~$\mu$m, corresponding to a minimum temporal delay step of $6.67$~fs. The time delay imparted by the delay stage is adjusted to reconcile the path lengths of the pump and probe beam paths such that the pump and imaging pulses arrive at the Kerr medium simultaneously, allowing careful overlap of the pulses in space and time. A neutral density (ND) filter is used to set the power of the pump beam and a half waveplate is used to rotate its polarization with respect to the imaging light. This adjustment is necessary to maximize the induced birefringence in the Kerr medium to fully utilize the transmission efficiency of the time-gate. The transmission of imaging light through the OKE gate is given by \cite{Alfano1984a}:
\[\frac{I_{t}}{I_{0}}=\sin^{2}\left(\frac{\Delta\phi}{2}\right)\sin^{2}(2\theta)\]
where $\Delta\phi$ is the phase shift experienced by the imaging pulse, and $\theta$ is the angle between the imaging and pump beam polarization states.

After interaction with the Kerr medium, the imaging beam path continues through a second dichroic mirror and band pass filter to eliminate the $800$~nm pump light, and through a bandpass filter designed to transmit the $400$~nm imaging light. The filtered imaging light then crosses a second polarizer aligned to block the original polarization of the imaging pulse (before interaction with the Kerr medium) to arrive at the detector forming a time-gated image.

Note that the pump beam $\lambda=800$~nm (derived directly from the laser beam) and probe beam $\lambda=400$~nm (frequency doubled using BBO crystal) are chosen in this fashion and not the other way round because the average power (or energy) required to introduce a significant OKE in the Kerr medium (CS$_2$) would not be enough due to the losses during SHG or otherwise an even higher power fs laser source would be required. Also, it has been observed that the high-power laser beam at $\lambda=400$~nm tends to damage the CS$_2$ by forming a brown precipitate over a very short duration of time. It is also for this reason that the CS$_2$ should not be placed at the Fourier plane of the lens $L_1$ in this dual-color configuration.

The detector used in the imaging measurements was an EM-CCD camera (Hamamatsu C9100-02). Supplementary intensity measurements and scans for system characterization were carried out using a high-resolution power meter (Ophir, Nova-II).

\begin{figure}[htbp]
\centering
\subfigure[For non-collinear or crossed-beam configuration]{\includegraphics[width=0.8\columnwidth]{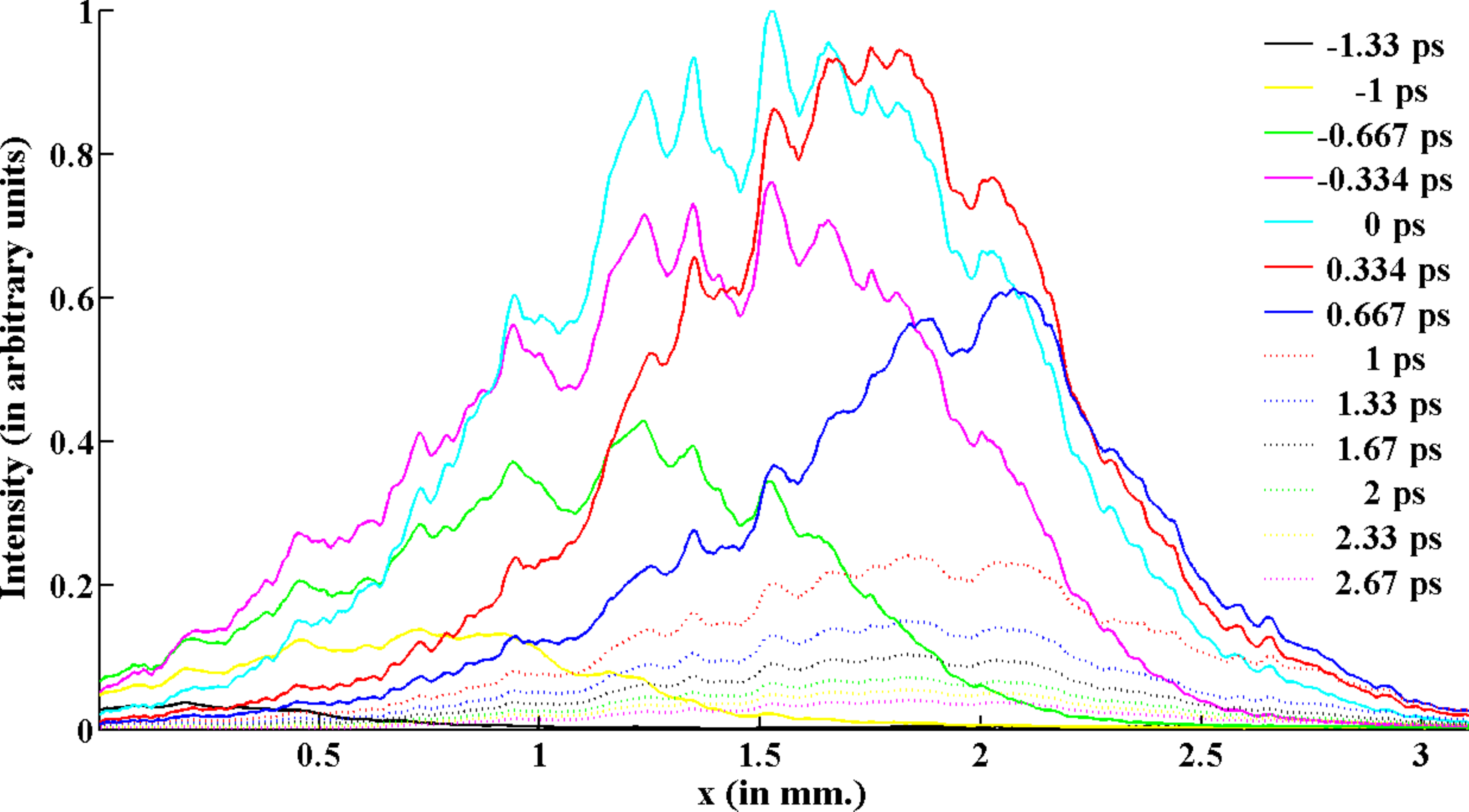}} \\
\subfigure[For collinear configuration]{\includegraphics[width=0.8\columnwidth]{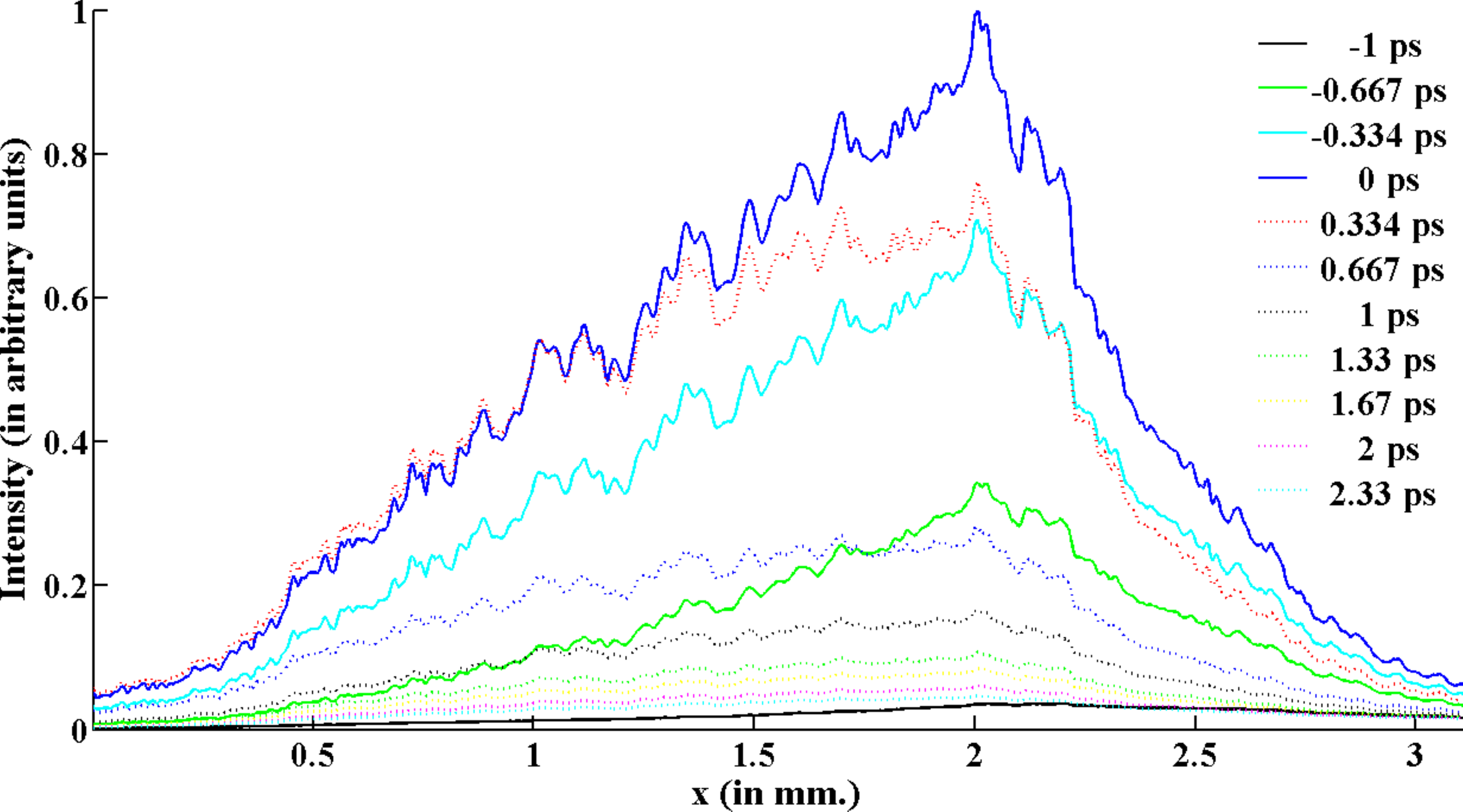}}
\caption{Average line profiles of the probe beam after passing through $1.0$~mm CS$_2$ optical time-gate for (a) non-collinear and (b) collinear incidence of the pump and probe beams at the Kerr medium for different delays between the two.}
\label{diff_coll}
\end{figure}
Most of the OKE time-gated imaging setups discussed in the literature employ a crossed-beam geometry in order to spatially separate the pump and image light \cite{Tan2010,Heisler2005,Tong2012,Idlahcen2009}. The spatial overlap of the pump beam on the probe beam covers an area inside the Kerr medium corresponding to a constant time. This solves the problem of mixing of spatial and temporal information carried by the probe as it passes through the optical gate. Hence, the output images for various delays between the pump and the probe pulses have spatial information pertaining to the same time event in this collinear approach. Figure~\ref{diff_coll} shows the average line profiles of the probe beam through the optical time-gate with non-collinear and collinear incidence of the incoming pump and probe beams at the Kerr medium for different delay between them. Since the overlap between the pump and probe pulses when incident collinearly is uniform in space, there is no shift in the intensity profile with delay in this case. This makes it more suitable for applications in ultrafast time gated imaging, since the information carried by the probe is not corrupted as it passes through the optical gate.
\pagebreak
\section{Results}
\subsection{Characterization of the optical time-gate}
\begin{figure}[htbp]
\centering
\includegraphics[width=0.8\columnwidth]{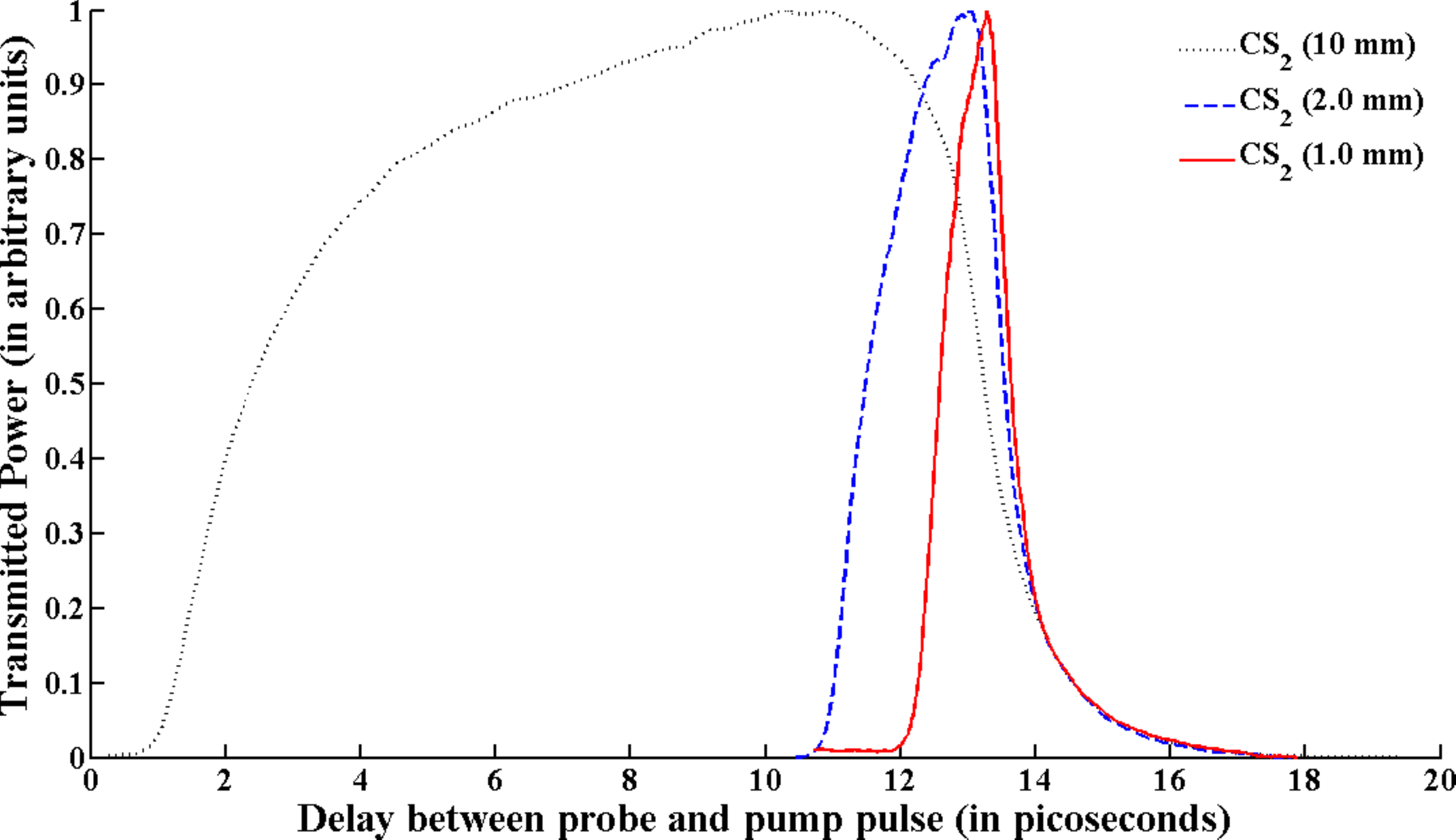}
\caption{Temporal profile of the dual wavelength, collinear OKE based time-gate with liquid CS$_2$ as Kerr medium for three different thicknesses of the CS$_2$ cell - $10$~mm, $2.0$~mm and $1.0$~mm. The average power of the pump beam for all these measurements was $0.34$~W.}
\label{CS2_temporal}
\end{figure}

The spatial resolutions and temporal profiles for the proposed OKE based time-gate presented above were measured with liquid carbon disulfide (CS$_2$) as Kerr medium known to produce measurable Kerr effect due to its high non-linear refractive index (n$_2$). Figure~\ref{CS2_temporal} shows the temporal profile of the optical time-gate for three different thicknesses of the CS$_2$ cell - 10~mm, 2.0~mm and 1.0~mm, without any object/sample placed in the probe beam's path. The gate duration (time for which the optical time-gate is open or allows a part of the probe pulse to pass through the crossed polarizer arrangement) is least in case of a 1.0~mm CS$_2$ cell.

The dependence of the optical gate duration on the thickness of the Kerr medium is not observed in classical single wavelength OKE-based time-gates. Indeed, in this case, the pump and the probe beams intersect within the Kerr medium, overlapping over a constant temporal width, in and out of the medium as they always travel with the same speeds. In the proposed dual wavelength configuration, since the wavelengths of both the beams are different ($\lambda=400$~nm for probe and $\lambda=800$~nm for pump), they propagate with different group velocities ($v_g$) inside the Kerr medium ($v^{probe}_g< v^{pump}_g$). So, even if a delay is imposed between these pulses outside the Kerr medium, the pump pulse catches up with the probe while traveling through the medium. An overlap occurs for a short duration and induces a polarization change of the probe beam, resulting in its transmission through the second polarizer, giving one point of the curves plotted in Fig.~\ref{CS2_temporal}. The range of possible overlapping is limited by the thickness of the Kerr medium. Nevertheless, a too thin Kerr medium leads to a poor efficiency of the optical time-gate. A more detailed discussion on the use of two wavelengths in optical gating may be found in ref. \cite{Idlahcen2009}. Figure~\ref{CS2_temporal} shows that a cell thickness of $1.0$~mm is a good compromise: it gives a temporal resolution of $1.04$~ps (full width at half maximum) and allows one to recover the relaxation time of CS$_2$.

\begin{figure}[htbp]
\centering
\subfigure[Measured and fitted ESFs, (in inset) slanted edge ($245$~$\mu$m glass fibre).]{\includegraphics[width=0.8\columnwidth]{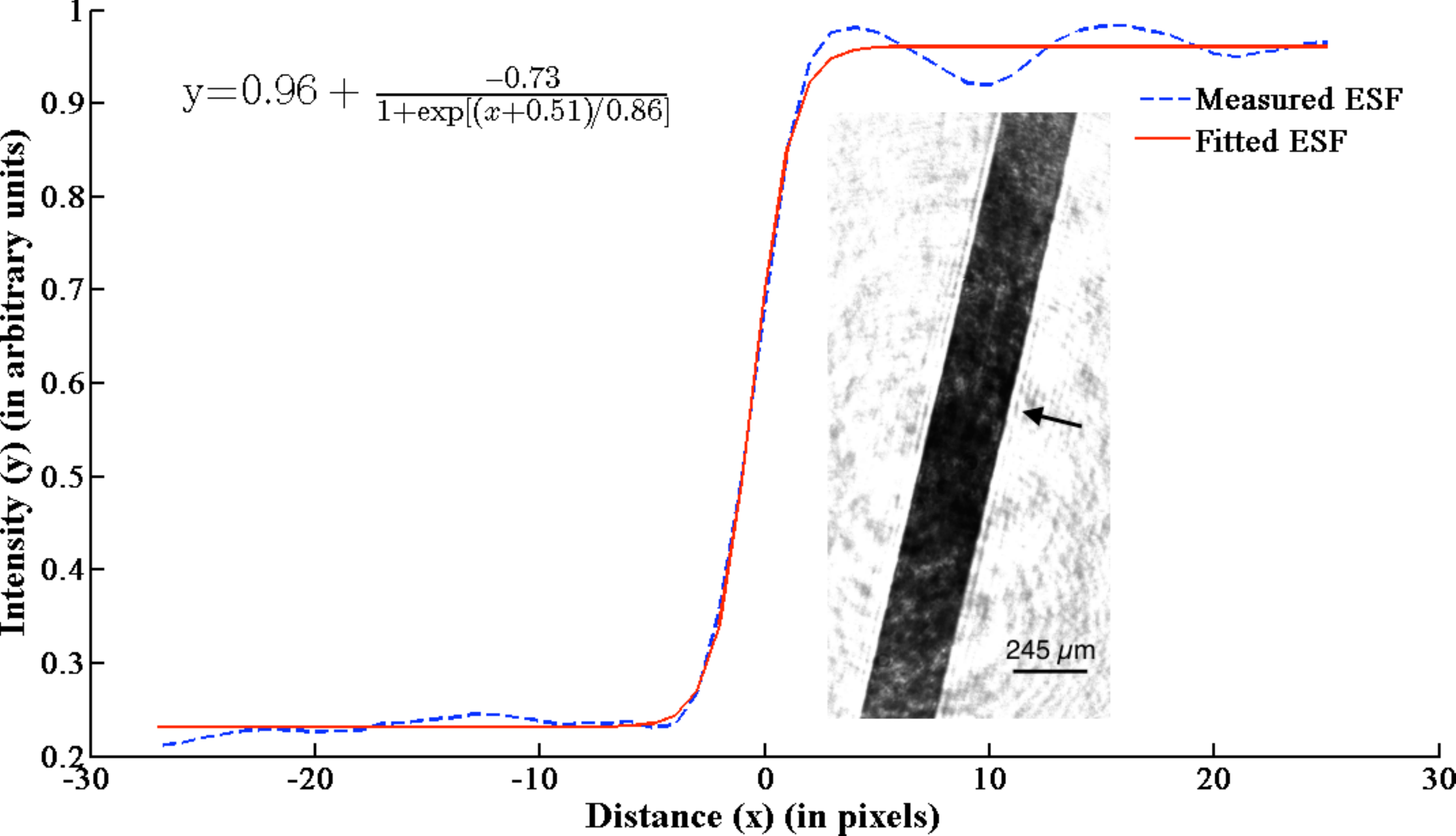}} \\
\subfigure[Normalized MTF vs spatial frequency ($\nu$)]{\includegraphics[width=0.8\columnwidth]{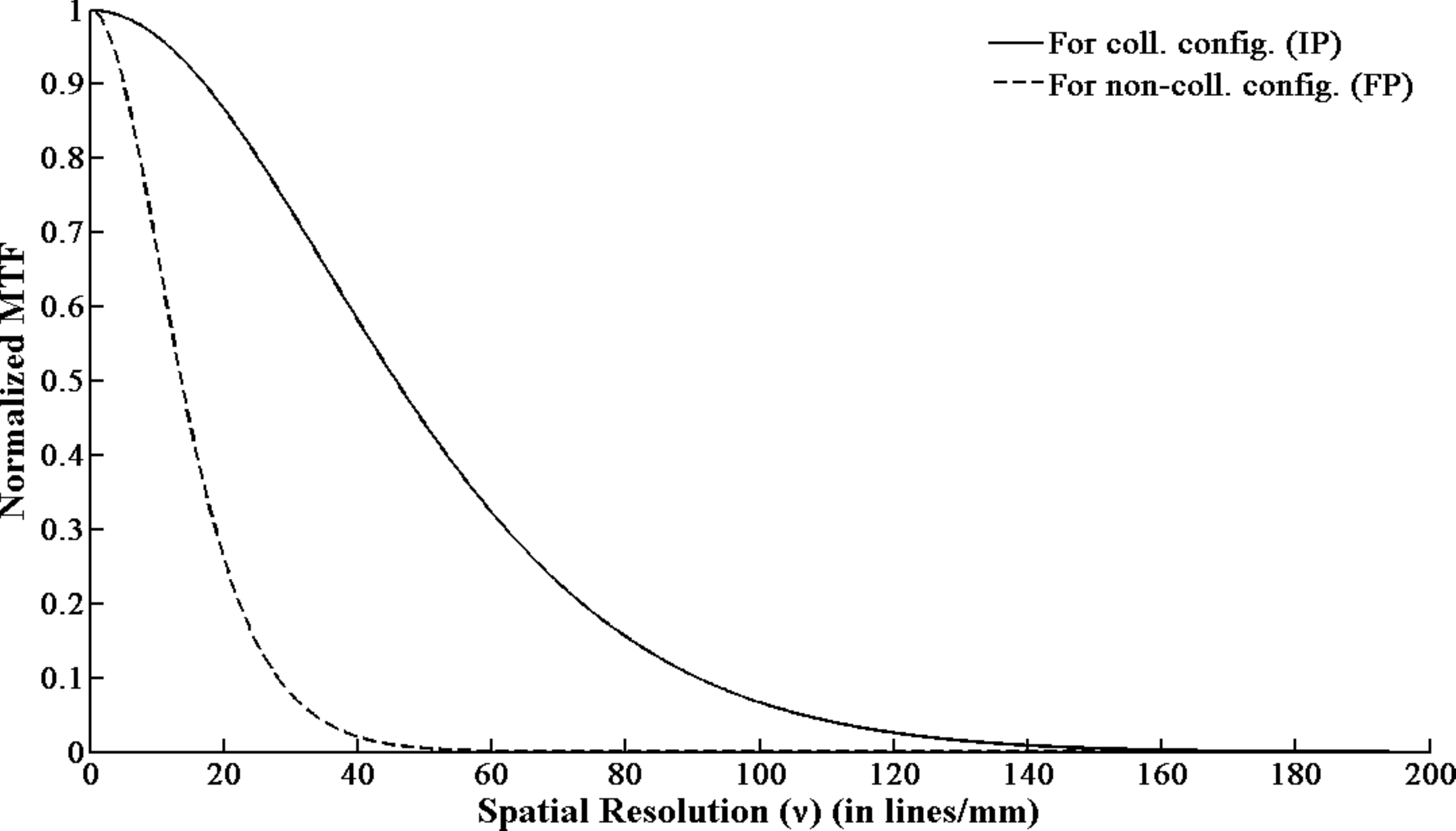}}
\caption{(a) Image of a slanted glass fibre ($245$~$\mu$m) obtained using collinear time-gate configuration with 1.0 mm thick CS$_2$ cell and the corresponding averaged edge-spread function (ESF) for the indicated edge in the image. (b) The normalized modulation transfer function (MTF) as a function of spatial frequency ($\nu$) (in lines/mm) calculated using the fitted ESFs for the collinear and crossed-beam configurations of the optical gate. Note that in the collinear configuration the Kerr medium (CS$_2$) was placed at the image plane (IP) of the lens $L_1$ whereas in the crossed-beam configuration it was placed at the Fourier plane (FP) of $L_1$.}
\label{mtf}
\end{figure}
Apart from the temporal resolution, another critical parameter for an optical gate is the quality of the measured images, and particularly spatial resolution, which may be estimated using the normalized modulation transfer function (MTF). The MTF for the present optical setup was measured using slanted-edge method (ISO 12233 standard for MTF measurement of electronic still-picture cameras \cite{iso}). The MTF and edge-spread function (ESF), $F(x)$ are related as follows,
\begin{equation} \label{mtfeq}
{\rm MTF}(\nu)=\int{\frac{dF(x)}{dx}e^{-i2\pi\nu x}dx}
\end{equation}
where $\nu$ is the spatial frequency in lines/mm, $x$ is the abscissa along an axis orthogonal to the slanted edge (or knife). In other words, MTF is the Fourier transform of the derivative of the ESF, which is also the line-spread function. It should be noted that while calculation of MTF from the over-sampled ESF data, the differentiation along the slanted edge is very sensitive to the high-frequency noise and in order to avoid amplifying this noise during the computation of the derivative, the averaged over-sampled ESF is fitted using the Logistic (Fermi) function $f(x)$ \cite{Li2009},
\begin{equation}
F(x)=d+f(x)=d+\frac{a}{1+\exp[(x-b)/c]}
\end{equation}
where $a$, $b$, $c$ and $d$ are the desired fit parameters.
Figure~\ref{mtf} shows the fitted ESF and normalized MTF calculated using eq.~\eqref{mtfeq} for dual-color, collinear configuration with the Kerr medium (CS$_2$) placed at the image plane of the lens $L_1$. For comparison Fig.~\ref{mtf}(b) also shows the normalized MTF for the traditional single-color, non-collinear configuration with CS$_2$ placed at the Fourier plane of $L_1$. The achievable imaging spatial resolution from the MTF curve (inverse of Nyquist frequency) with CS$_2$ as the Kerr medium placed at the image plane of $L_1$ for collinear configuration was found to be $95.5$~lines/mm and for the non-collinear configuration with the same Kerr medium placed at the Fourier plane of $L_1$ it was found to be $37.6$~lines/mm.
\subsection{Examples of application}
\begin{figure}[htbp]
\centering
\subfigure[Ballistic image]{\includegraphics[width=0.4\columnwidth]{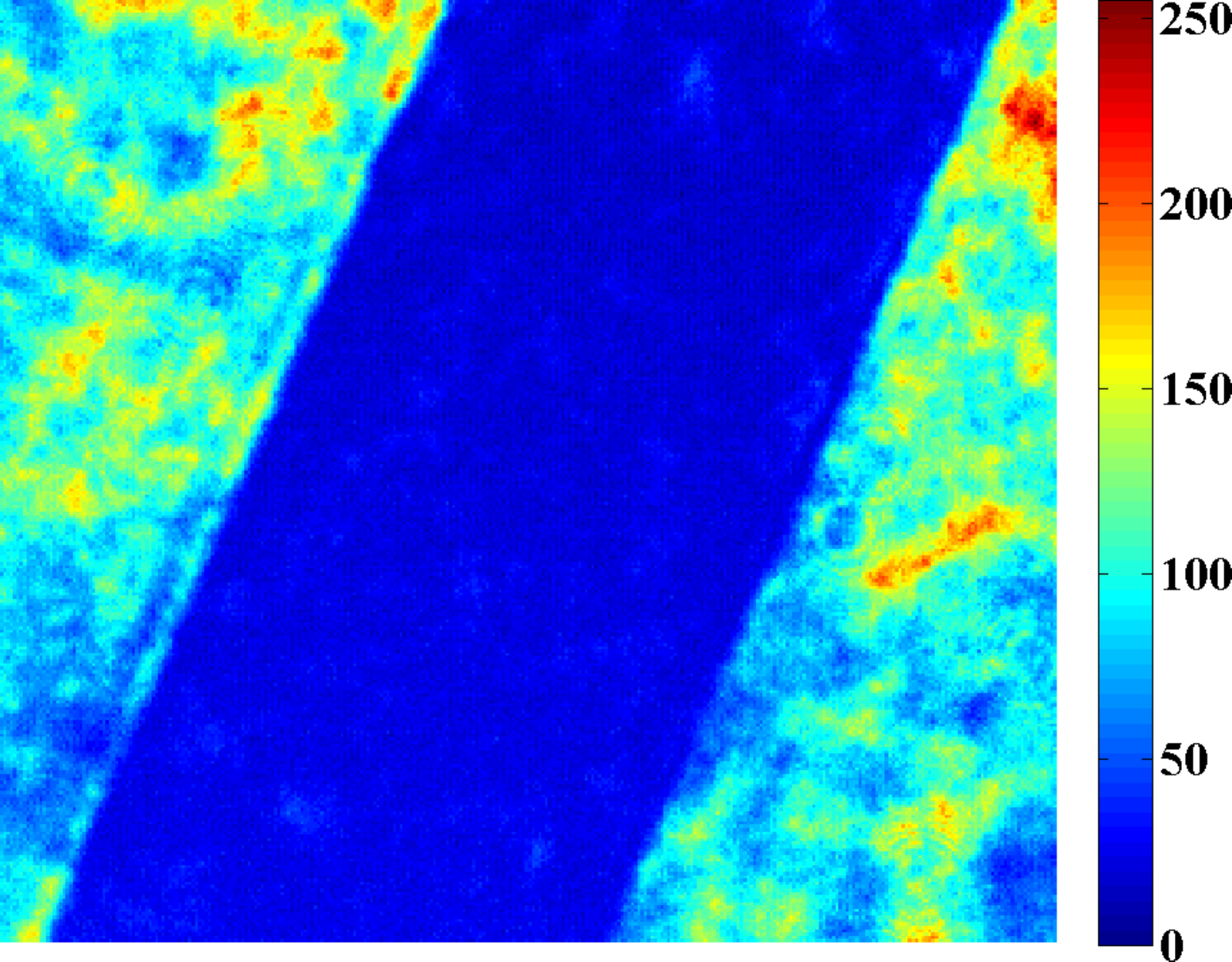}}
\quad
\subfigure[Refraction image]{\includegraphics[width=0.4\columnwidth]{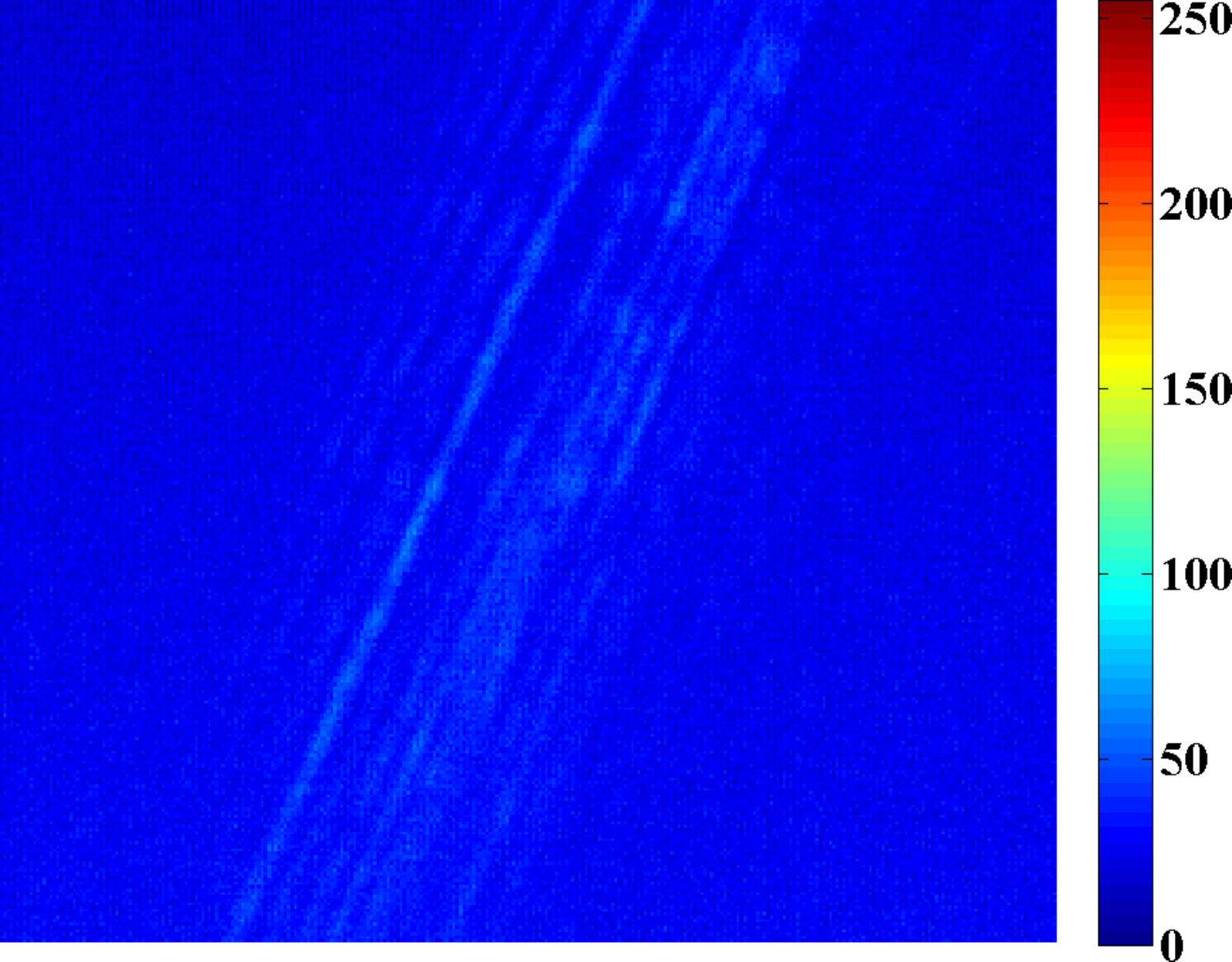}}
\caption{Ballistic and refraction images of a $342$ $\mu$m micro-structured optical fibre (LMA-25) obtained using $1.0$ mm CS$_2$ optical gate. The hidden microstructures are visible in the refraction image. Average pump power $0.34$ W.}
\label{struct_fibre_bal_ref}
\end{figure}

Beyond the development and characterization of the proposed optical time-gate setup, an application of the technique for ballistic imaging is shown for different objects. Figure~\ref{struct_fibre_bal_ref} shows images of a micro-structured optical fibre LMA-25 from NKT Photonics with an overall average diameter of $342$ $\mu$m (with coating) obtained using $1.0$~mm CS$_2$ optical gate. In Fig.~\ref{struct_fibre_bal_ref}(a), the delay line is adjusted to select only ballistic light, i.e. light which has not interacted with the object. When the delay between the probe and pump is increased, ballistic light disappears and the hidden microstructures inside the optical fibre as in Fig.~\ref{struct_fibre_bal_ref}(b) are revealed by the refracted light, though the intensity of the refracted light is very low as compared to the ballistic light.

\begin{figure}[htbp]
\centering
\subfigure[Ballistic image]{\includegraphics[width=0.8\columnwidth]{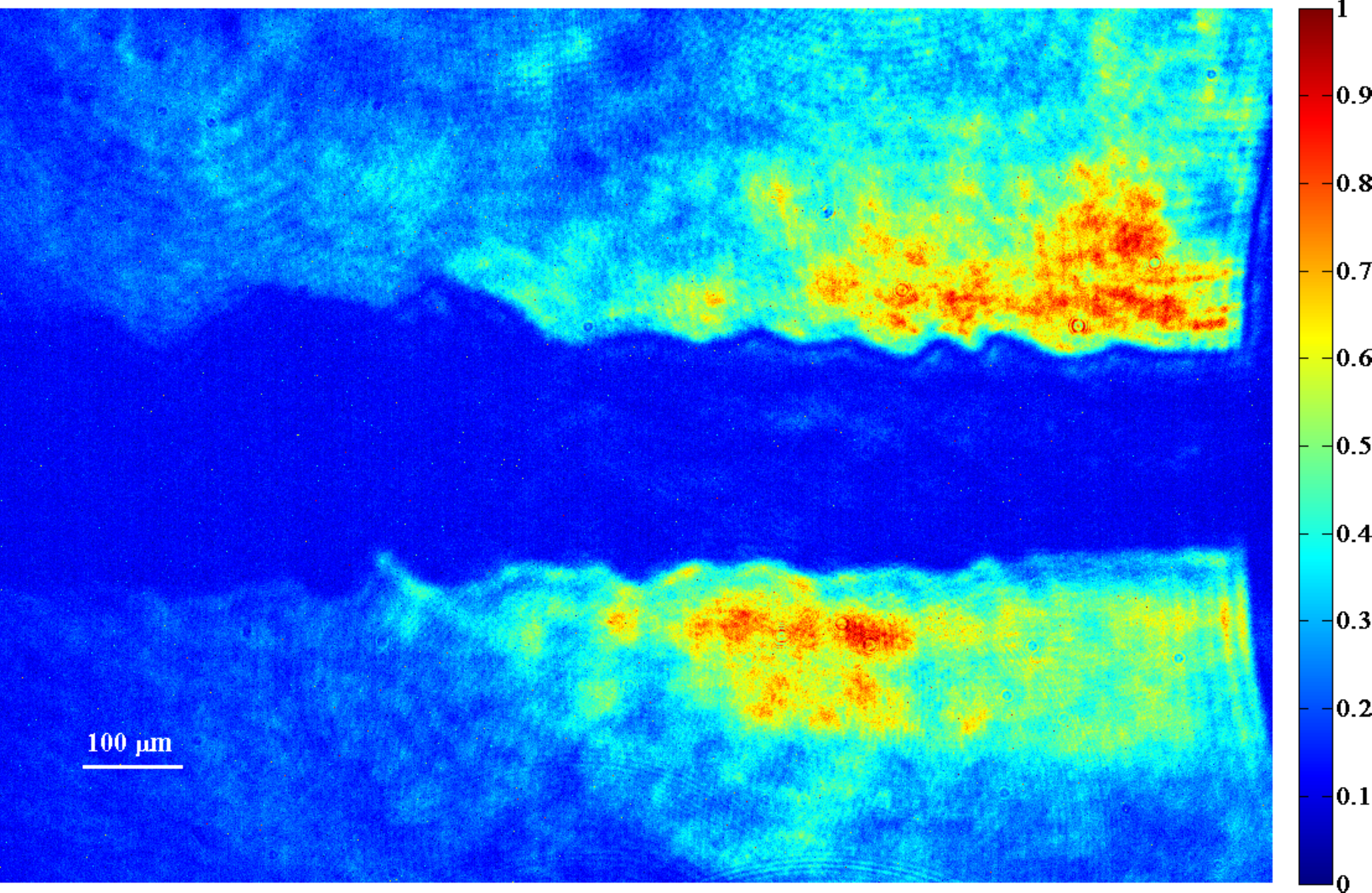}}\\
\subfigure[Refraction image]{\includegraphics[width=0.8\columnwidth]{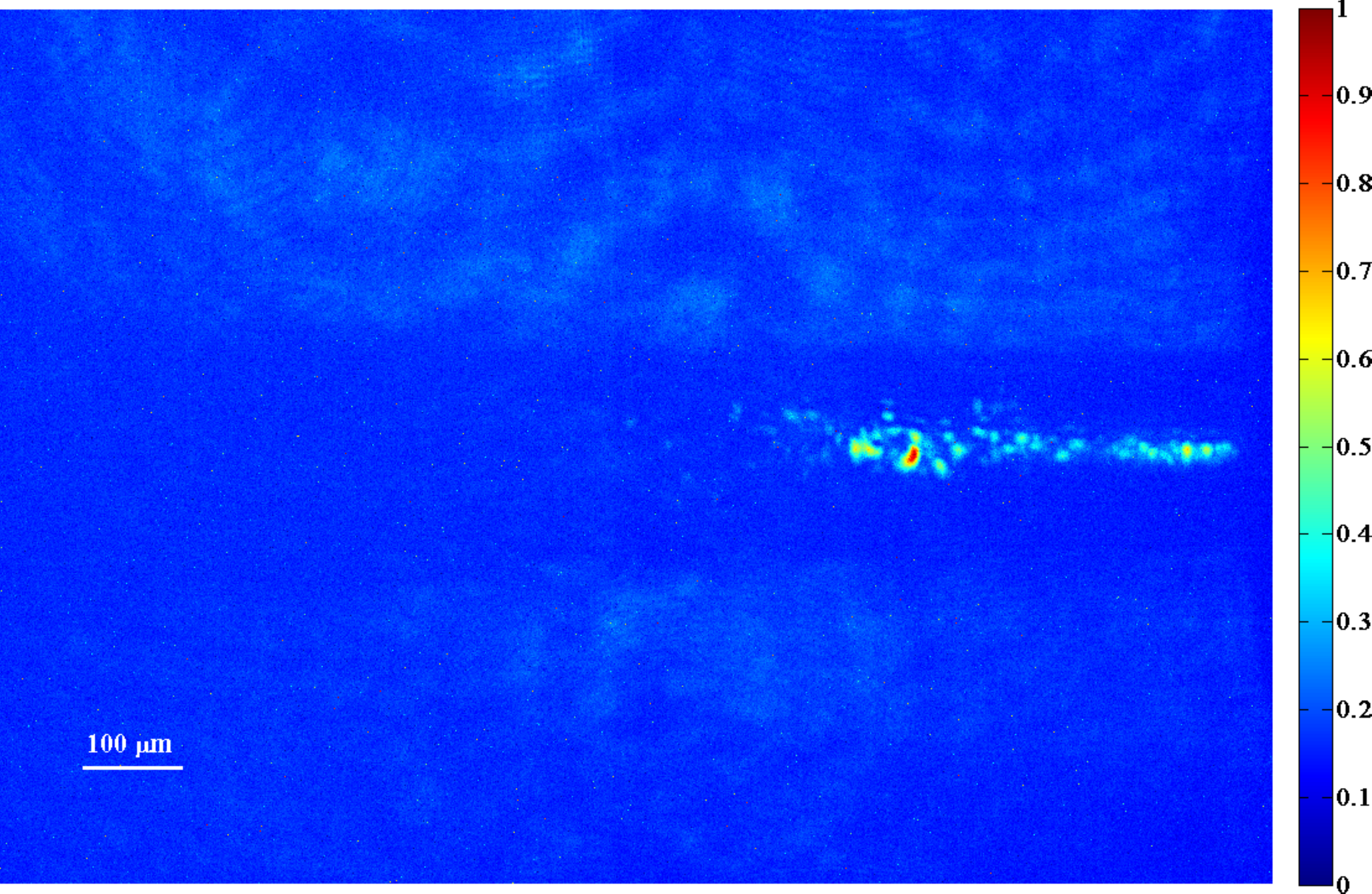}}
\caption{Ballistic and refraction images of fuel spray close to the injector (Bosch, single orifice $\phi=185$~$\mu$m) separated using $1.0$ mm CS$_2$ optical time-gate.}
\label{spray_bal_ref}
\end{figure}
Figure~\ref{spray_bal_ref} shows an application of the $1.0$~mm CS$_2$ optical time-gate to a liquid spray, produced by a single orifice injector (diameter $\phi=185$~$\mu$m). The injection pressure was $45$~MPa, and the resulting velocity about $130$~m/s. The fluid used in the measurements was a calibration oil (Shell NormaFluid, ISO 4113) with thermophysical properties similar to that of diesel fuel, and a refractive index of $n=1.46$. The images were taken close to the injector and the first millimeter from the nozzle is visible. Again, the delay between the pump and the probe was adjusted so that detection time coincides with ballistic light Fig.~\ref{spray_bal_ref}(a). When the delay was increased, ballistic light disappeared and a small fluctuating bright line appeared at the center of the jet, corresponding to the refracted light (Fig.~\ref{spray_bal_ref}(b)), showing that at a few hundreds of micrometers from the nozzle, the liquid jet shape is still close to a cylinder.

\begin{figure}[htbp]
\centering
\includegraphics[width=0.7\columnwidth]{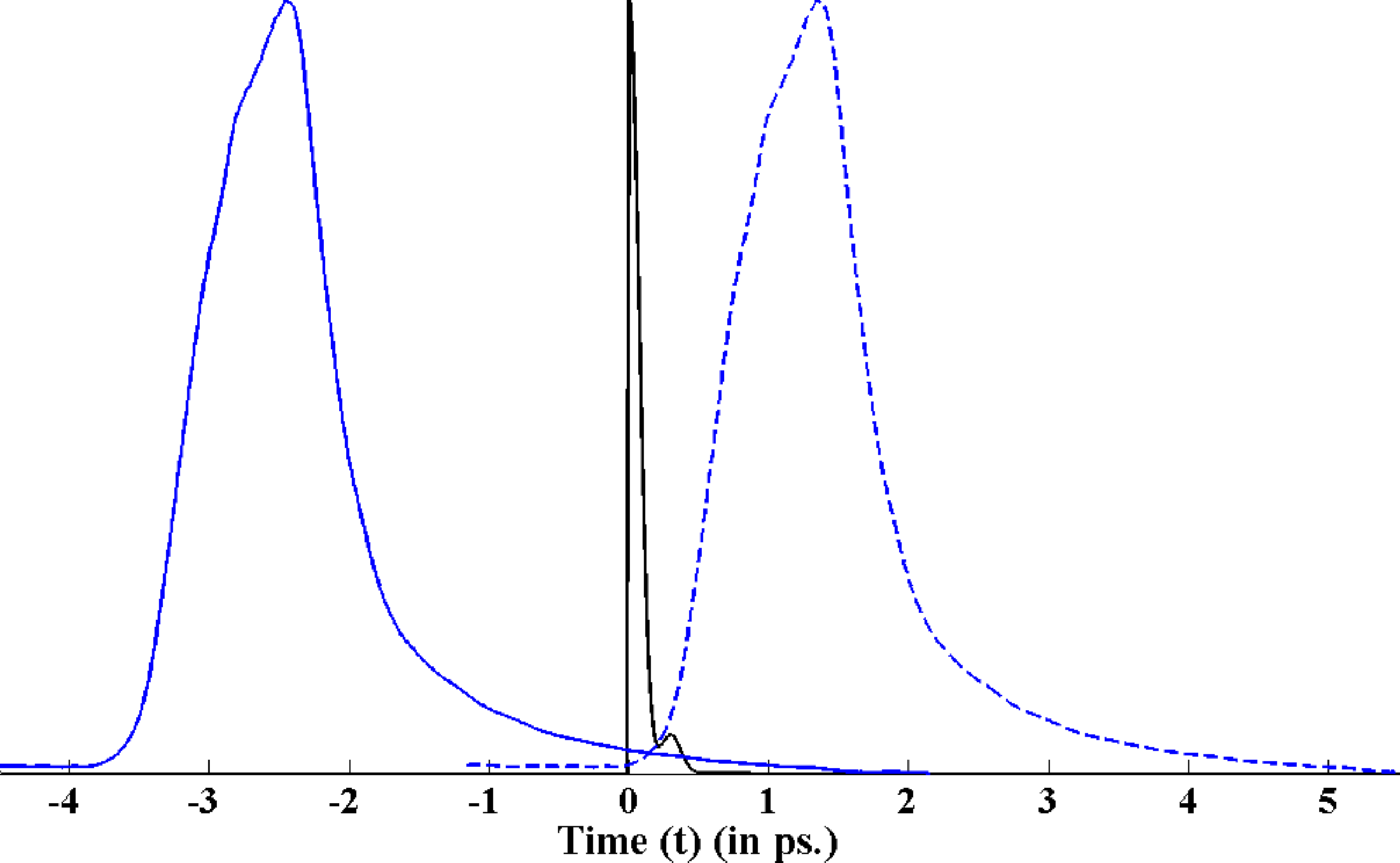}
\caption{Schematic to show the ballistic and the refracted light signals through a $185$~$\mu$m thick medium with the refractive index of $1.46$ (black, solid line curve). The blue, solid and dashed curves shows the transmission through the $1.0$~mm CS$_2$ optical time-gate (gate duration $\sim1.0$~ps) for two different delay configurations, separating ballistic and refraction light signals.}
\label{schematic}
\end{figure}
It should be noticed that the gate duration for $1.0$~mm CS$_2$ based time-gate is about $1$~ps, and seems not short enough to separate the ballistic and the refracted light signals: the $185$~$\mu$m diameter corresponds to a $300$~fs time delay between them. However, the separation of the refracted and the ballistic light signals is possible even by a time gate of much larger duration, by imposing an appropriate delay between the pump and the probe pulses (Fig.~\ref{schematic}). Since the ballistic light is much more intense compared to the refracted light, for an appropriate delay (blue, solid line curve of Fig.~\ref{schematic}) even a small transmission through the time-gate is sufficient to produce a high contrast ballistic image at the detector. The contribution of the refracted light becomes negligible at this delay. And then by preciously increasing the delay between the pump and the probe pulses (blue, dashed line curve of Fig.~\ref{schematic}), it is possible to obtain a similar image with contribution mainly from the refracted light.

\section{Conclusions}
A collinear, dual wavelength OKE based time-gate has been proposed to overcome the drawbacks of the classical single wavelength, crossed-beam arrangement. The idea of using different wavelengths for pump and probe beams reduces the noise due to the scattering of the high power pump beam towards the detector. It opens the possibility to combine the pump and the probe pulses into a collinear beam, removing the mixing of spatial and temporal information on the images. The achievable spatial resolution estimated from MTF calculation was found to be $95.5$ lines/mm with Kerr medium (CS$_2$) placed at the image plane of the lens $L_1$. The time resolution was about $1$~ps, limited by the relaxation time of CS$_2$. The developed optical time-gate was found to be capable of separating ballistic and refracted light from a static object (a microscopic optical fiber) and a dynamic one (a liquid spray).

\section*{Acknowledgments}
This work was supported by NADIA-Bio program, with funding from the French Government and the Haute-Normandie region, in the framework of the Moveo cluster ("private cars and public transport for man and his environment").
\end{document}